\documentclass{elsart}
\usepackage{graphicx}
\usepackage{bm} 
\begin{document}
\bibliographystyle{num}

\begin{frontmatter}
\title{\bf Nuclear matter with three-body forces from self-consistent spectral 
calculations\thanksref{grant}}
\thanks[grant]{Supported by 
Polish Ministry of Science and Higher Education under
grant N202~034~32/0918}  
\author[INP]{Vittorio Som\`a}, and 
\author[INP,UR]{Piotr Bo\.zek},
\address[INP]{The H. Niewodnicza\'nski Institute of Nuclear Physics, \\
         Polish Academy of Sciences, PL-31342 Krak\'ow, Poland}
\address[UR]{Institute of Physics, Rzesz\'ow University, PL-35959 Rzesz\'ow, Poland}

\begin{abstract} 
We calculate  the equation of state of nuclear matter
in the self-consistent T-matrix scheme including  three-body nuclear interactions. We study the effect of the three-body force on the self-energies and spectral functions of nucleons in medium.
\end{abstract}


\end{frontmatter}

\vspace{-7mm} PACS:  21.65.+f, 24.10.Cn, 25.30.-c

The extrapolation of the energy per particle in  dense nuclear systems 
from finite nuclei to infinite nuclear matter and to  neutron matter 
in neutron stars
involves theoretical estimates. Most of the approaches are using free 
nucleon-nucleon potentials when treating  systems with many 
particles. Obviously 
a number of approximations must be done to obtain a result.
Properties of dense nuclear matter have been estimated with
 a variety of different schemes~:
Brueckner-Hartree-Fock (BHF) calculations \cite{brugam,jlm,baldonstar,lombardonstar}, variational calculations \cite{vcs1,vcs2,fabrocini1,fabrocini,v18} and
self-consistent T-matrix calculations 
\cite{Bozek:2002em,Bozek:2002tz,Dewulf:2002gi,Dewulf:2003nj,Frick:2003sd}.
The central issue of these studies is the
 calculation of the binding energy of cold symmetric and neutron rich nuclear matter.  It has been realized that a
 realistic description of the nuclear matter
 at saturation density and beyond cannot neglect the presence of
 three-body forces between nucleons. Variational 
and BHF calculations that take the three-body forces into account 
reproduce the empirical saturation point density and the binding
 energy in symmetric nuclear matter 
\cite{Baldo:2001mw,3holes,Baldo:2007wm,vcs1,vcs2,Li:2006gr}.
The spectral T-matrix method 
 using in medium dressed nucleon propagators and the
  self-consistent T-matrix is well suited for the calculation of 
 single-particle properties of nucleons in the 
 medium \cite{Bozek:2002em}, the effective 
scattering \cite{Dickhoff:1999yi} and pairing correlations 
\cite{Bozek:1999rv,Bozek:2001nx}.  
One important feature of the scheme
 is the automatic fulfillment of thermodynamic consistency relations 
\cite{Bozek:2001tz}.
The equation of state of symmetric and
neutron matter in the T-matrix approach have been evaluated as well
 \cite{Dewulf:2002gi,Bozek:2002ry,Bozek:2002tz,Dewulf:2003nj},
 however only two-body 
interactions have been included up to now. 
The resulting equation of state is 
similar to the 
one obtained by other methods with the same interactions. 
In this letter we 
present first results for the self-consistent T-matrix 
approximation with 
phenomenological three-body interactions taken into account. After reproducing 
the properties of the  symmetric nuclear matter  
around the
 empirical saturation point we calculate the modified in-medium properties
 of dressed nucleons at several densities.

The in-medium self-consistent T-matrix 
\cite{KadanoffBaym,Danielewicz:1982kk,Alm:1996ps} 
is defined by the two-body potential
 $V$ 
as 
\begin{equation}
T=V+VGGT
\end{equation}
which denotes the  resummation of the two-nucleon ladder diagrams.
The in-medium nucleon propagator
\begin{equation}
G=\frac{1}{\omega - p^2/2m - \Sigma}
\end{equation}
is dressed by the self-consistently calculated self-energy
\begin{equation}i\Sigma=Tr[TG] \ . \end{equation}
The numerical calculations are performed in the real-time formalism for the 
finite-temperature Green's functions \cite{Bozek:2002em}.
The spectral function for dressed nucleons is obtained from the retarded
 self-energy (with energy and momentum arguments explicitly written)
\begin{equation}
A(p,\omega)=\frac{-2\mbox{Im}\Sigma(p,\omega)}{(\omega-p^2/2m-\mbox{Re}\Sigma(p,\omega))^2
+\mbox{Im}\Sigma(p,\omega)^2} \ .
\end{equation}
At each density $\rho$ the above set of equations is iterated until 
convergence and the Fermi energy $\mu$ is adjusted to fulfill the constraint
\begin{equation}
\rho=\int_{-\infty}^\mu \frac{d\omega}{2\pi} \int \frac{d^3p}{(2\pi)^3}
 A(p,\omega) \ .
\end{equation}
The binding energy can be calculated from the Galitskii-Koltun's sum rule
\begin{equation}
\frac{E}{N}=\frac{1}{2\rho}\int \frac{d\omega}{2\pi} \int \frac{d^3p}{(2\pi)^3}
 (\omega+p^2/2m)A(p,\omega) \ ,
\end{equation}
a formula that works for Hamiltonians with 
two-body interactions. In the general case one should calculate the diagrams 
corresponding to the expectation value of the Hamiltonian $<H>$
 \cite{Soma:2006zx}.

Several parameterizations of the nuclear two and three-body 
interactions are used in nuclear matter and finite nuclei calculations. 
Since the short range behavior of the nuclear force is not precisely known, 
differences in parameterization of the nucleon-nucleon interaction at 
high momenta can be compensated by differences in the associated 
three-body term \cite{Grange:1989nx,Zuo:2002sf}. Explicit calculations
 using relativistic BHF formalism require
 a different (if any)  three-body force. Still other three-body 
interactions are needed when using renormalized effective two-body potentials 
\cite{Bogner:2005sn}.
In the following we use a simple phenomenological way of taking the
 three-body interaction into account \cite{old3b} and 
for the two-body potential we take the CD-Bonn interaction.
The three body term motivated by the two-pion exchange process 
has  the form of an additional density dependent two-body interaction
\begin{figure}[tb]
\begin{center}
\includegraphics[width=12cm]{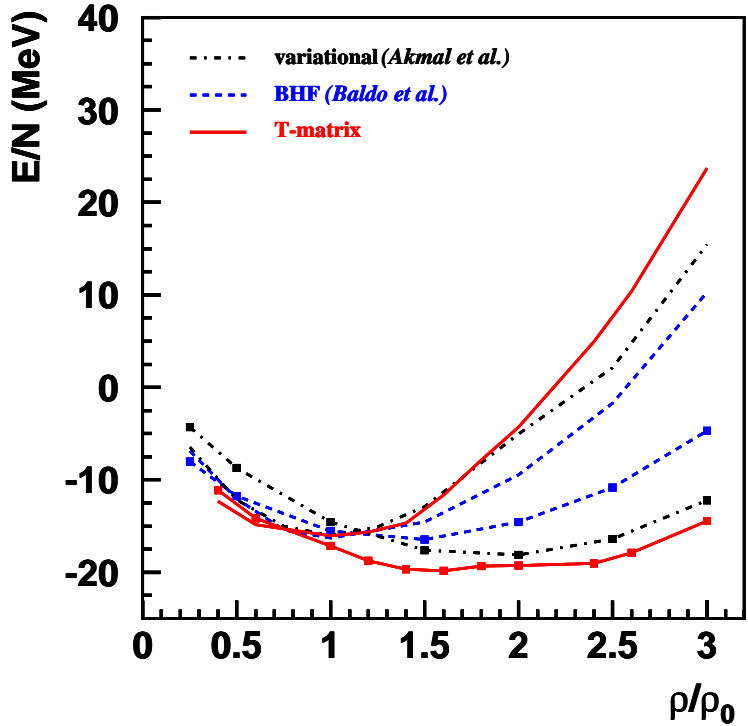}
\end{center}
\caption{The equation of state for symmetric nuclear matter. The solid
 line represents the result of the T-matrix calculation, the dashed line
 the BHF result \cite{Baldo:2007wm} and the dashed line the variational
 result \cite{vcs2}. The lines with symbols denote the results of  
calculations including only the two-body force.}
\label{binden}
\end{figure}
\begin{equation}
\label{2bint}
V(r)=I^c T^2(r) \left(e^{-\gamma_1 \rho}-1\right)
\end{equation}
where
\begin{equation}
T(r)=\left( 1+\frac{3}{\eta r}+\frac{3}{\eta^2 r^2}\right)\frac{e^{-\eta r}}
{\eta r}
\left( 1 - e^{-c r^2}\right)^2 , 
\end{equation}
$\eta=0.7$fm$^{-1}$, $c=2$fm$^{-2}$, $I^c=-5.7$MeV, $\gamma_1=0.15$fm$^3$.
The density dependent part is short range and repulsive and leads to a 
stiffening of the equation of state.
We perform iterative self-consistent calculation
 of the dressed propagators and the in-medium
$T$ matrix with such density dependent interactions at several
 densities between $\rho=0.6\rho_0$ and $\rho=3\rho_0$, where
 the saturation density is $\rho_0=0.17$fm$^{-3}$. Additionally 
a phenomenological 
attractive mean field energy is taken in the form \cite{old3b}
\begin{equation}
E_{TNA}(\rho)/N=3\gamma_2 \rho^2 e^{-\gamma_3\rho} \ , 
\end{equation}
where $\gamma_2= -260 $MeV fm$^6$ and $\gamma_3= 11 $fm$^3$.
With the density dependent two-body interaction (\ref{2bint}) 
the Galitskii-Koltun's sum rule gives the same binding energy
 as the expectation value of the Hamiltonian. For the chosen interaction 
the difference shows up only in the term $E_{TNA}$, that should be 
added to the energy but the corresponding shift in the Fermi energy is 
$\delta (\rho E_{TNA}(\rho)/N)/\delta \rho$.

The  binding energy in symmetric nuclear matter is shown in
Fig.~\ref{binden} 
and compared to results of variational and BHF calculations 
with two-body and three-body forces \cite{Baldo:2007wm,vcs2}.
 We observe that the inclusion of density 
dependent forces suffices to reproduce the empirical saturation point of 
nuclear matter,  although our equation of state is slightly stiffer with 
the parameters of the short range density dependent interaction taken directly 
from 
\cite{old3b}.
As expected, in order to get realistic binding energies at higher densities
 three-body forces must be included in the calculation.
\begin{figure}[tb]
\begin{center}
\includegraphics[width=12cm]{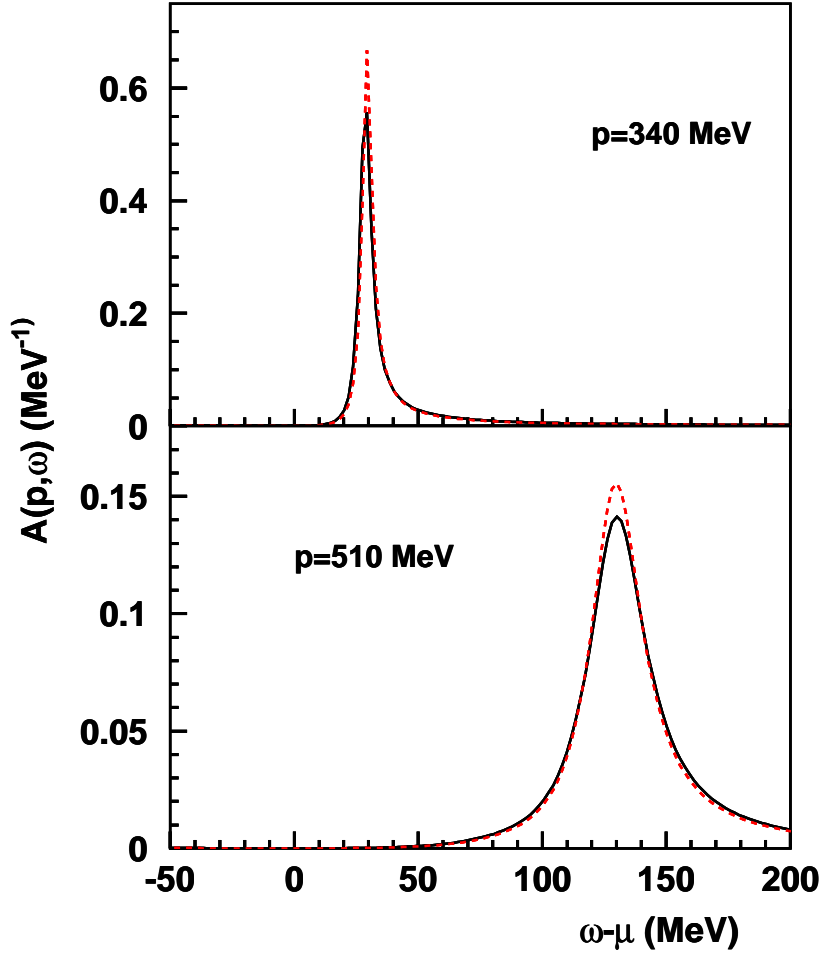}
\end{center}
\caption{The spectral function for dressed nucleons at saturation density. 
The dashed line is the result with the three-body force,
 the solid line represents 
the result of the calculation the with two-body interactions only.}
\label{spectr}
\end{figure}
The equation of state fixes the strength of the additional three body terms. 

Within the T-matrix scheme we explore the consequences of the
 additional interaction terms 
on the single-particle properties. In Fig. \ref{spectr} is shown the spectral 
function $A(p,\omega)$ at $\rho_0$. We see that the modification of the force 
changes the spectral function. This indicates that besides the equation
 of state the spectral function is sensitive to the chosen nuclear interaction.
 The binding energy is not sensitive enough to constraint the short range part 
of the nuclear interaction. The nuclear spectral function when compared 
to experimental results  \cite{Rohe:2004dz} gives insight 
into the nuclear two and three-body interactions in nuclei. In a future
 publication we shall present the results of an investigation of the 
sensitivity 
of the proton spectral functions on the density and isospin dependence 
of the three-body force. Within the parametrization of the three-body terms 
taken in this letter, the spectral functions shows a stronger quasi-particle
 peak, which  is a consequence of a reduced scattering at the Fermi surface.

\begin{figure}[tb]
\begin{center}
\includegraphics[width=12cm]{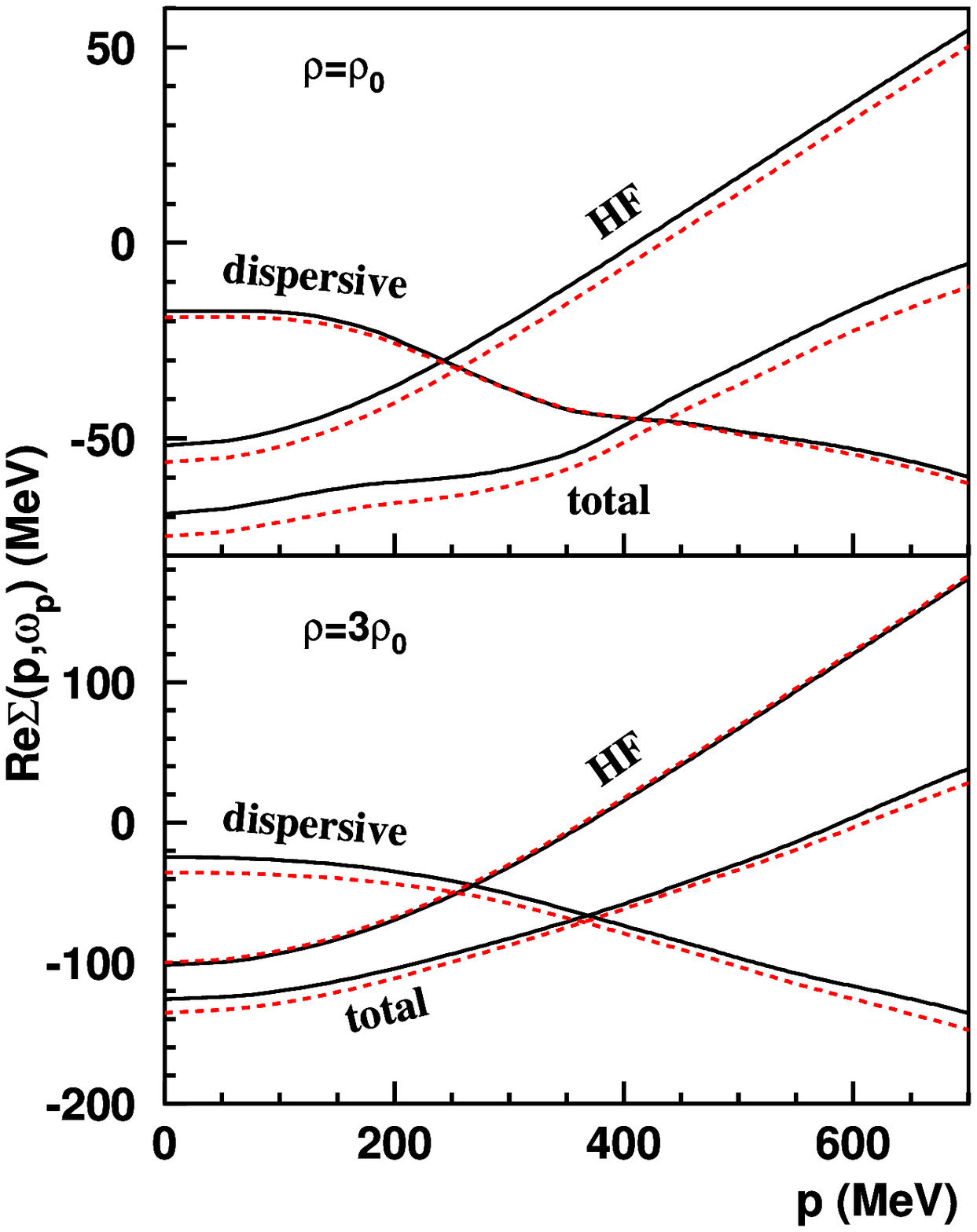}
\end{center}
\caption{The real part of the self-energy at the quasiparticle pole
 $\mbox{Re}\Sigma(p,\omega_p)$, at the saturation density $\rho_0$ 
(upper panel) and at $3\rho_0$ (lower panel). We show the Hartree-Fock 
contribution, the dispersive part, and the sum  (Eq. \ref{selfen}).
The dashed lines are the results including three-body forces 
and the solid lines represent the energies obtained with two-body forces only.}
\label{realself}
\end{figure}
It is interesting to analyze the real-part of the self-energy 
$\mbox{Re}\Sigma(p,\omega_p)$ at the 
quasi-particle pole ($\omega_p=p^2/2m + \mbox{Re}\Sigma(p,\omega_p)$). 
It  is  the in medium potential 
felt by the nucleon. 
The self-energy is given by dispersion relation
\begin{eqnarray}
\mbox{Re}\Sigma(p,\omega)&=& \Sigma_{HF}(p)+\int \frac{d\omega^{'}}{\pi}\frac{\mbox{Im}
\Sigma(p,\omega^{'})}{\omega^{'}-\omega}
 \nonumber \\
& = & \Sigma_{HF}(p) + \Sigma_{disp}(p,\omega) 
\label{selfen}
\end{eqnarray}
It is a sum of a dispersive self-energy $\Sigma_{disp}$ and the
 mean-field self-energy
$\Sigma_{HF}(p)$.      Different parameterizations of the nuclear interactions 
yield different  Hartree-Fock and dispersive self-energies, hard-core 
potentials give less attractive mean-field. However the total self-energy 
$\mbox{Re}\Sigma(p,\omega)$ is similar
 \cite{Bozek:2002tz}\footnote{Interactions yielding similar binding energies
must give similar single-particle energies by thermodynamic consistency}.
At saturation density the dispersive part is not modified much 
by the three-body forces, only the Hartree-Fock energy is lowered. 
On the other hand at $\rho=3\rho_0$, 
where the scattering is more important, the
 dispersive part of the potential is lowered when including three-body forces, 
but the mean-field is similar. Depending on the density, 
the shift of the single-particle energy caused by the three-body forces
 manifests itself in a different way. In all cases the total self-energy 
$\mbox{Re}\Sigma(p,\omega_p)$ is more attractive when three-body 
forces are taken 
into  account.

We calculate for the first time
 the properties of nuclear matter in the self-consistent
 T-matrix approximation with a phenomenological density 
dependent three-body term in the nucleon-nucleon interaction.
The additional terms in the interaction and in the mean-field energy 
allow to reproduce the empirical saturation point of symmetric nuclear matter.
We find a slightly stiffer equation of state at densities above $2 \rho_0$
than other approaches (BHF and variational). We calculate also 
the spectral function and the in-medium nuclear self-energy. The spectral 
function shows a stronger quasi-particle peak and the potential 
for the nucleons 
is more attractive when three-body forces are
 included. The equation of state and the nucleon potential are not directly 
sensitive to the details of the short range part of the nuclear interaction.
On the other hand the proton
spectral function at high energies  could be a 
probe of short range nuclear correlations. 
For the interaction  studied in this paper we find less scattering 
when density dependent three body terms are included.

 \bibliography{../mojbib}

\end{document}